# Doppler peaks from active perturbations

Joao Magueijo[1], Andreas Albrecht[2], David Coulson[3], Pedro Ferreira[4]

[1] Mullard Radio Astronomy Observatory, Cavendish Laboratory, Madingley Road, Cambridge, CB3 0HE, U.K.
and
Department of Applied Mathematics and Theoretical Physics, University of Cambridge, Cambridge CB3 9EW, U.K.
[2] Blackett Laboratory, Imperial College, Prince Consort Road London SW7 2BZ U.K.
[3] D. Rittenhouse Laboratory, University of Pennsylvania, Philadelphia, PA, 19104-6396
[4] Center for Particle Astrophysics, University of California Berkeley, CA 94720-7304

**Abstract**

We examine how the qualitative structure of the Doppler peaks in the angular power spectrum of the cosmic microwave anisotropy depends on the fundamental nature of the perturbations which produced them. The formalism of Hu and Sugiyama is extended to treat models with cosmic defects. We discuss how perturbations can be "active" or "passive" and "incoherent" or "coherent", and show how causality and scale invariance play rather different roles in these various cases. We find that the existence of secondary Doppler peaks and the rough placing of the primary peak unambiguously reflect these basic properties.

The cosmic microwave background (CMB) promises to become one of the most successful bridges between theory and experiment in cosmology. As the body of experimental data continues to grow[1] theorists are evaluating the impact of this data on the two major paradigms for cosmic structure formation: inflation [2], and topological defects [3]. One of the important theoretical tools is the formalism of Hu and Sugiyama (HS)[4] which we extend here to accommodate topological defects.

The so-called Doppler peaks, in particular, have attracted great interest. They consist of a system of oscillations, known to be present for most inflationary models, in the CMB angular power spectrum $C^l$ at $100 < l < 1500$. The peaks' height and position can be used to fix with some accuracy combinations of parameters left free in inflationary models [5]. Progress on defect Doppler peak predictions has been slow (see however [6, 7, 8]). It was suggested in [8] that, regardless of the remaining quantitative uncertainties, one could expect dramatic qualitative differences between defect and inflationary Doppler peaks. More concretely, it was pointed out that the non-existence of secondary peaks is a robust feature of some defect theories resulting from the different role played by randomness and causality in these theories. In this *Letter* we elaborate on how general this feature is, and pin down its controlling factors.

The idea is to focus on the basic assumptions of inflationary and defect theories, isolate the seminal contrasting properties, and perform a qualitative analysis of the structure of the Doppler peaks purely based on these properties. Inflationary fluctuations were produced at a remote epoch, and were driven far outside the hubble radius by inflation. The evolution of these fluctuations is linear (until gravitational collapse becomes non-linear at late times), and we call these fluctuations "passive". Also, because all scales observed today have been in causal contact since the onset of inflation, causality does not strongly constrain the fluctuations which result. In contrast, defect fluctuations are continuously seeded by defect evolution, which is a non-linear process. We therefore say these are "active" perturbations. Also, the constraints imposed by causality on defect formation and evolution are much greater than than those placed on inflationary perturbations.

The notions of scale invariance and causality have different implications in these two types of theory. A scale invariant gauge-invariant potential $\Phi$ (the Newtonian potential on subhorizon scales) with dimensions $L^{3/2}$ has a power spectrum $P(\Phi) = \langle |\Phi_{\mathbf{k}}|^2 \rangle \propto k^{-3}$ in passive theories (the Harrison-Zeldovich spectrum). In active theories the most general counterpart to the Harrison-Zeldovich spectrum is $P(\Phi) = \eta^3 F_\Phi(k\eta)$. Moreover, active perturbations are constrained by causality, in the form of integral constraints [9, 10], such as those written in terms of the gauge dependent energy-momentum pseudo-tensor of [11, 12]. In [13] we show that the density subject to the integral constraint can be written in the gauge-invariant form $\mathcal{U} = a^2\rho\Delta_T + \rho^s + 3hv^s$. Here, $a$ is the scale factor, $h = \dot{a}/a$, $\rho$ ($\Delta_T$) is the total matter density (density contrast) and the scalar defect stress-energy tensor is given by $\Theta_{00} = \rho^s$, $\Theta_{0i} = k_i v^s$, and $\Theta_{ij} = p^s\delta_{ij} + (k_i k_j - k^2\delta_{ij}/3)\Pi^s$. Then following [14], on superhorizon scales $P(\mathcal{U}) \propto k^4$ for active perturbations. The Einstein equations for the scalar gauge-invariant potentials, $\Phi$ and $\Psi$ are [15, 16]

$$k^2\Phi = 4\pi\left(a^2\rho\Delta_T + \rho^s + 3hv^s\right), \qquad (1)$$

$$\Phi + \Psi = -8\pi\left(a^2\frac{p\Pi}{k^2} + \Pi^s\right). \qquad (2)$$

Since isotropy requires $\Pi^s$ and $p\Pi/k^2$ (where $p\Pi/k^2$ is simply related to the quadrupole of the photon and neutrino fluctuations) to be constant for small $k$, the Einstein equations imply that scaling active perturbations produce scaling gauge-invariant potentials, which must be white-noise on large scales. In particular $P(\Psi - \Phi) = F(k\eta)\eta^3$, with $F(0)$ a non-zero constant. For most realistic defects $x^4 F(x)$ will then have a single peak, located at a value of $x$ close to the location of the peak of the defect compensated structure function [17]. We will see that the place and thickness of the peak in $x^4 F(x)$ are deciding features for the Doppler peaks induced by active perturbations.

Active perturbations may also differ from inflation in the way "chance" comes into the theory. Randomness occurs in inflation only when the initial conditions are set up. Time evolution is linear and deterministic, and may be found by evolving all variables from an initial value equal to the square root of their initial variances. By squaring the result one obtains the variables' variances at any time. Formally this results from unequal time correlators of the form

$$\langle \Phi(\vec{k},\eta)\Phi(\vec{k}',\eta')\rangle = \delta(\vec{k}-\vec{k}')\sigma(\Phi(k,\eta))\sigma(\Phi(k,\eta')), \tag{3}$$

with $\sigma(\cdot) = \sqrt{P(\cdot)}$. In defect models however, randomness may intervene in the time evolution as well as the initial conditions. Although deterministic in principle, the defect network evolves as a result of a complicated non-linear process. If there is strong non-linearity, a given mode will be "driven" by interactions with the other modes in a way which will force all different-time correlators to zero on a time scale characterized by the "coherence time" $\tau_c(k,\eta)$. Physically this means that one has to perform a new "random" draw after each coherence time in order to construct a defect history [8]. The counterpart to (3) for incoherent perturbations is

$$\langle \Phi(\vec{k},\eta)\Phi(\vec{k}',\eta')\rangle = \delta(\vec{k}-\vec{k}')P(\Phi(k,\eta),\eta'-\eta) . \tag{4}$$

For $|\eta' - \eta| \equiv |\Delta\eta| > \tau_c(k,\eta)$ we have $P(\Phi(k,\eta),\Delta\eta) = 0$. When convolving $P(\Phi(k,\eta),\Delta\eta)$ with functions which vary slowly at the scale of $\tau_c(k,\eta)$ we may implement an approximation where

$$\langle \Phi(\vec{k},\eta)\Phi(\vec{k}',\eta')\rangle = \delta(\vec{k}-\vec{k}')\delta(\eta-\eta')P^r(\Phi(k,\eta)), \tag{5}$$

in which

$$P^r(\Phi(k,\eta)) = \int d\Delta\eta P(\Phi(k,\eta),\Delta\eta) \tag{6}$$

is the time-integrated power spectrum [17]. We shall label as coherent and incoherent the perturbations satisfying (3) and (5) respectively. This feature changes the way the average $C^l$ are computed, resulting in a striking qualitative difference in the structure of Doppler peaks. We expect that equation (3) and (5) will be only rough approximations for some defect cases but still allow some intuition to be gained.

A large class of theories is embraced by combinations of the two concepts just introduced. Inflationary perturbations are passive coherent perturbations. Defect perturbations are active perturbations more or less incoherent depending on the defect. We will submit this class of theories to the computational machinery of Hu and Sugiyama [4], which was initially tailored for passive coherent perturbations. We consider the limit where $\Omega = 1$ and $\Omega_b = 0$. Generalization is straightforward from HS. Whereas primordial terms are dominant for passive perturbations, for active perturbations one may drop all but the convolution terms. The radiation brightness multipoles at epoch $\eta_0$ ($\eta_0 > \eta^*$, the epoch of decoupling) are then given by

$$\Theta^l(k,\eta_0) = \int_0^{\eta^*} d\eta\, k(\Phi - \Psi)(D^l(\eta) + V^l(\eta))$$
$$+ [(\Psi - \Phi)A^l](\eta^*) + \int_{\eta^*}^{\eta_0} d\eta\, (\dot\Psi - \dot\Phi)A^l(\eta) \quad (7)$$

with $A^l(\eta) = (2l+1)j_l(k(\eta_0 - \eta))$ and the projected monopole and dipole contributions given by

$$D^l(k,\eta) = \frac{(2l+1)j_l(\sqrt{3}\Delta x)}{\sqrt{3}} \sin \Delta x,$$
$$V^l(k,\eta) = ((l+1)j_{l+1}(\sqrt{3}\Delta x) - lj_{l-1}(\sqrt{3}\Delta x)) \cos \Delta x,$$

where $\Delta x = k(\eta^* - \eta)/\sqrt{3}$. The average $C^l$ are computed in HS assuming (3). For incoherent perturbations one has instead

$$C^l = \pi^2 \int dk \int_0^{\eta^*} d\eta\, k^4 P^r(\Phi - \Psi)\left(\frac{D^l + V^l}{2l+1}\right)^2$$
$$+ \int dk\, k^2 P(\Psi - \Phi)(\eta^*)\left(\frac{A^l(\eta^*)}{2l+1}\right)^2 + \int dk\, k^3 \frac{[V^l A^l \sigma^r(\Phi G)\sigma^r(\Psi - \Phi)](\eta^*)}{(2l+1)^2}$$
$$+ \pi^2 \int dk \int_{\eta^*}^{\eta_0} d\eta\, k^2 P^r(\dot\Psi - \dot\Phi)\left(\frac{A^l(\eta)}{2l+1}\right)^2 \quad (8)$$

resulting from squaring and averaging (7) using statistics as in (5). These generalizations, together with the active perturbations' gauge-invariant potentials, allow a systematic extension of the HS formalism to the defect case.

We first undertake a preliminary analysis by examining the power spectrum of the radiation energy density (the monopole term $\Theta_0 + \Psi$) at last scattering. This should mimic the Doppler peaks' qualitative structure. As in [4] we have for adiabatic and isocurvature passive fluctuations:

$$k^3 P(\Theta_0 + \Psi)(\eta^*) \approx k^3 P(\Theta_0 + \Psi)(0) \cos^2 x^*, \quad (9)$$
$$k^3 P(\Theta_0 + \Psi)(\eta^*) \approx k^3 P(\Theta_0 + \Psi)(0) \sin^2 x^*, \quad (10)$$

in which $x^* = k\eta^*/\sqrt{3}$. The peaks of the spectrum are at the scales $x_m^* = m\pi$ for adiabatic, and $x_m^* = (m - 1/2)\pi$ for isocurvature perturbations. These correspond roughly to the angular scale $l_m \approx x_m^* \eta_0$. For coherent active perturbations $k^3 P(\Theta_0 + \Psi)(\eta^*)$ is approximately

$$\left(\frac{1}{\sqrt{3}}\int_0^{\sqrt{3}x^*} dx\, x^{\frac{3}{2}} F^{\frac{1}{2}}(x) \sin\left(x^* - \frac{x}{\sqrt{3}}\right) - x^{*\frac{3}{2}} F^{\frac{1}{2}}(x^*)\right)^2,$$

whereas for incoherent perturbations one has

$$x^{*3} F(x^*) + \frac{1}{3}\int_0^{\sqrt{3}x^*} dx\, x^4 F(x) \sin^2\left(x^* - \frac{x}{\sqrt{3}}\right),$$

where $P(\Phi - \Psi) = \eta^3 F(x)$ (or $P^r(\Phi - \Psi) = \eta^4 F(x)$), with $x = k\eta$. These show that the position and structure of active perturbations Doppler peaks result from a combination of the issue of coherence and the details of the potential structure function $F(x)$. If $x^4 F(x)$ has a sufficiently thin peak at $x = x_c \equiv 2\pi\eta/\xi_c$ (where $\xi_c$ is approximately the coherence length of the defect) then the monopole spectrum peaks will be at $x_m^* = (m - 1/2)\pi + x_c/\sqrt{3}$ for both coherent and incoherent fluctuations. Then active perturbations just apply a phase shift of value $x_c/\sqrt{3} - \pi/2$ to an adiabatic type of spectrum.

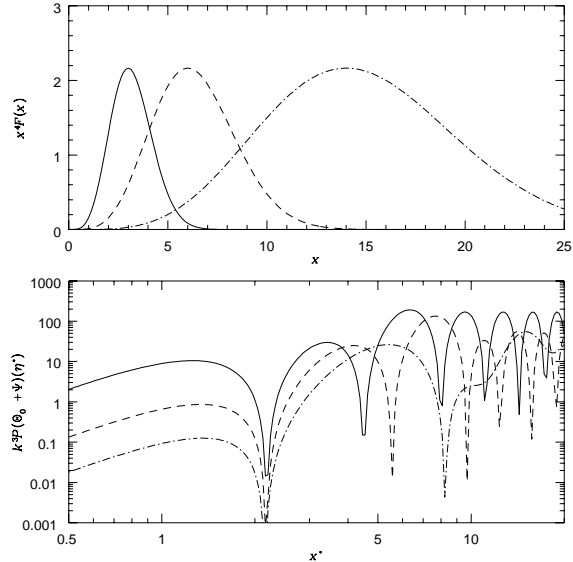

Figure 1: Three potential structure functions $F(x)x^4$ and their corresponding spectra $k^3 P(\Theta_0 + \Psi)(\eta^*)$ assuming coherence. By increasing $x_c$ one moves from adiabatic peaks (line) to isocurvature peaks (dash). For larger $x_c$ the secondary peaks come out significantly distorted (point-dash).

For $x_c \approx 2.7$ (unrealistic because it is very close to the smallest turnover point allowed by causality[10]) the monopole peaks are at the adiabatic positions. For all other causal active perturbations the peaks are shifted to smaller scales. For $x_c \approx 3.4$ they are out of phase with the adiabatic peaks (as in [6]). For $x_c > 5.5$ the peaks start only in the adiabatic secondary peaks region. For standard values of $\Omega_b$ and $h$ these three cases would place the main "Doppler peak" at $l \approx 230$, $350$, and $500$, respectively. Therefore the placing of the peaks is *not* a generic feature of active fluctuations. Active perturbations simply add an extra parameter on which the Doppler peaks position is strongly dependent. In general we should expect that for the same $\Omega$, $\Omega_b$, and $h$, active perturbations will take the predicted CDM adiabatic peak position $l$ to $l + \eta_0(x_c/\sqrt{3} - \pi/2)$. The secondary peaks' separation is not changed, in a first approximation. This is to be contrasted with non-flat inflationary models where $C_l(\Omega = 1)$ is taken into $C_{l\Omega^{1/2}}$. Thus it should be possible to distinguish between low $\Omega$ CDM and $\Omega = 1$ high-$x_c$ defects.

For large $x_c$ the peak in $x^4 F(x)$ can never be thin. Then each mode is active for several expansion times, bringing coherence into play. Qualitative changes come about in the secondary peaks, but our conclusions relating to the primary peak still hold. In Fig. 1 we consider coherent perturbations with realistic structure functions. One may obtain passive type of peaks at adiabatic and isocurvature positions. For $x_c > 5$ there are strong distortions. One must however realize that effective coherence for large $x_c$ requires actual coherence over several (more than $x_c$) expansion times, perhaps an unreasonable demand.

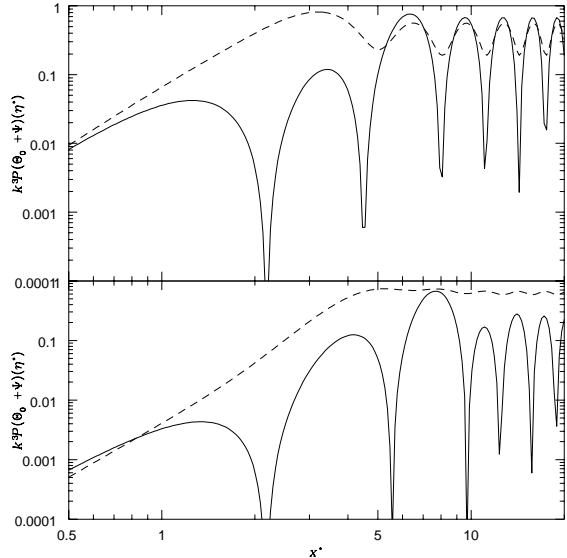

Figure 2: The $k^3 P(\Theta_0 + \Psi)(\eta^*)$ spectrum for coherent (line) and incoherent (dashed) active perturbations with the first two structure functions used in Fig. 1. One may obtain (softer) secondary oscillations at the adiabatic position for incoherent perturbations. As the spectrum shifts to the right (larger $x_c$) the secondary oscillations disappear very quickly.

The situation for incoherent oscillations is illustrated in Fig. 2. Although there are never true zeros in $P(\Theta_0 + \Psi)(x^*)$ it is possible to obtain significant oscillations if the main peak is at the adiabatic position. However these disappear very quickly as the main peak approaches the isocurvature position. Hence, large $x_c$ defects can always be expected to produce exotic $C^l$ spectra, as suggested in [8]. Coherence intervenes in deciding how large $x_c$ must be for this to happen, as well as what type of novelty is introduced.

Besides this qualitative general analysis, the extended HS formalism allows for an approximate solution (5-10%) for the $C^l$'s of any particular defect model. The calculation errors are in practice dominated by uncertainties in the defect stress-energy tensor. We illustrate the procedure with the example of cosmic strings. For these we use the incoherent form with $P^r(\rho^s) = 1/(1+2(\beta x)^2)$ (from [17]). We consider the two cases $\beta = 1$ and $\beta = .3$ similar to the X and I models in [17]. We consider only scalar contributions. We assume that the defect variables are subject to equations of state of the form $p^s = \gamma(x)\rho^s$, $\Pi^s = \eta^2 \gamma_s(x)\rho^s$, and $v^s = \eta \gamma_v(x)\rho^s$. Energy conservation at small $x$ requires that $3\gamma(0) = (1/2\alpha) - 1$ and $\gamma_v(0) = (1-2\alpha)/(3\alpha(4\alpha+1))$, with $\alpha = \eta h$. We make use of a string simulation to determine the large $x$ behavior[13]. We find, with large uncertainties, that $x\gamma_v(x) = s \approx .1 - .3$, and $x^2\gamma^s(x) = \sigma \approx .4 - .55$. We interpolate between the $x \gg 1$ and $x \ll 1$ behaviour. We set $\Pi = 0$ and assume that $a^2 \rho \Delta_T$ is subdominant except for the compensation. This consists of a white noise large-scale tail in the spectrum of $a^2 \rho \Delta_T$ present in order to cancel the white-noise tail in $\rho^s + 3hv^s$ and ensure that $P(a^2 \rho \Delta_T + \rho^s + 3hv^s)$ goes like $k^4$.

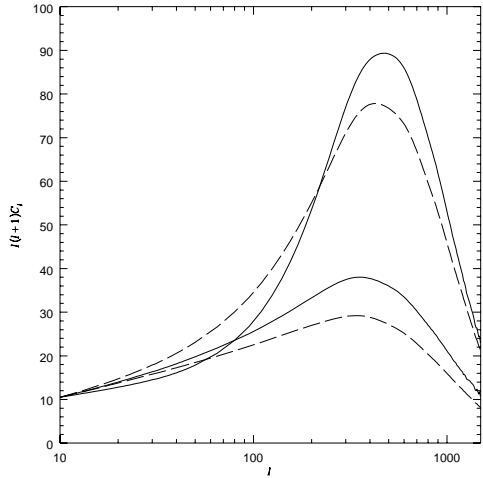

Figure 3: The $C^l$ spectrum for I (dash) and X (line) cosmic strings. The top lines use $s = .1, \sigma = .4$ for X strings and $s = .15, \sigma = .45$ for I strings. The bottom lines both use $s = .2, \sigma = .5$. We have assumed $\Omega = 1$, $h = .5$, and $\Omega_b = 0.05$.

We take the compensation into account by hand, setting $a^2 \rho \Delta_T + \rho^s + 3hv^s = \gamma_c(\rho^s + 3hv^s)$ with $\gamma_c = 1/(1 + (\alpha/x)^2)$. We fix the compensation scale at $\alpha = 2\pi$. Using (1) and (2) we finally obtain the required cosmic strings potential structure functions to be inserted in the HS formalism as modified for incoherent perturbations. The results are plotted in Fig. 3. The Sachs-Wolfe plateau exhibits a "running" tilt ranging from $n \approx 1.4$ before $l = 10$ to $n \approx 1.2$ at $30 - 40$. There is a single Doppler bump located at $l \approx 400 - 600$ These features are remarkably robust against uncertainties in the equations of state. The ratio between the peak and the plateau heights, on the other hand, can change by as much as an order of magnitude.

In general, we find that features of the $C^l$ spectrum suggested by the monopole spectrum at decoupling are confirmed. Generic defects place the main peak to the right of the CDM adiabatic peak. Coherent defects exhibit shifted CDM-type secondary oscillations up to the isocurvature positions (which are easily distinguished from the shifts associated with varying $\Omega$). From then on coherent defects show a peculiar type of secondary oscillations. Incoherent defects erase the secondary oscillations if the main peak is placed on or to the right of the isocurvature position. Thus the most dramatic effects occur for large $x_c$ defects (such as cosmic strings) where the $C^l$ spectrum shape at $100 < l < 1500$ is radically different according to the active/passive, and coherent/incoherent nature of the perturbations. The signature becomes progressively less prominent as $x_c$ is pushed to the lower limit imposed by causality.

ACKNOWLEDGMENTS: We acknowledge support from St.John's College, Cambridge (J.M.), PPARC (A.A.), DOE Grant DOE-EY-76-C-02-3071 (D.C.), and the Center for Particle Astrophysics, a NSF Science and Technology Center at UC Berkeley, under Cooperative Agreement No. AST 9120005 (P.F.),